\documentclass[pdftex,12pt,
epjc3]{svjour3}

\usepackage{bm}
\usepackage{hyperref}
\usepackage{mathrsfs}
\usepackage{xcolor,color,graphicx,graphics}
\usepackage[all]{xy}
\usepackage{latexsym,amssymb,amsmath,amsfonts} 
\usepackage[english]{babel} 
\usepackage[OT1]{fontenc}
\usepackage{makeidx}
\usepackage{hyperref}
\usepackage{color,graphicx,graphics,wrapfig,epsf}

\usepackage{newtxtext,newtxmath}

\usepackage[T1]{fontenc}

\begin{document}

\title{Application of Machine Learning Methods for Detecting Atypical Structures in Astronomical Maps}

\author{I.A. Karkin\thanksref{e1,addr1} \and  A.A. Kirillov\thanksref{e2,addr1}        \and
        E.P. Savelova\thanksref{e3,addr1}    
}

\thankstext{e1}{e-mail: karkinia@yandex.ru}
\thankstext{e2}{e-mail: ka98@mail.ru}
\thankstext{e3}{e-mail: sep$\_$ 22.12.79@inbox.ru}
\institute{Bauman Moscow State Technical University, Moscow,  105005, Russian Federation \label{addr1}
}

\date{
}

\maketitle

\begin{abstract}
The paper explores the  use of various machine learning methods to search for heterogeneous or atypical structures on astronomical maps. The study was conducted on the maps of the cosmic microwave background radiation from the Planck mission obtained at various frequencies. The algorithm used found a number of atypical anomalous structures in the actual maps of the Planck mission. This paper details the machine learning model used and the algorithm for detecting anomalous structures. A map of the position of such objects has been compiled. The results were compared with known astrophysical processes or objects. Future research involves expanding the dataset and applying various algorithms to improve the detection and classification of outliers.
\end{abstract}




\section{Introduction}

Significant progress in astronomy has been made in recent years due to the introduction of machine learning methods for analyzing observational data. The enormous amount of data generated by telescopes and other instruments complicates the manual search and identification of both interesting and unusual structures. Machine learning algorithms can effectively analyze large datasets in high-dimensional parameter spaces and identify patterns that would be difficult for humans to detect  \cite{1}. 

The Planck Space Observatory, launched by the European Space Agency and NASA in 2009, was designed to study the cosmic microwave background (CMB) radiation, the oldest light in the universe. CMB maps contain a wealth of valuable information about the early universe and its large-scale structure, making the mission extremely important for cosmology. The mission aimed to map the Big Bang's CMB with improved sensitivity and resolution and to test theories about the universe's origin and evolution. However, extracting and analyzing this information is an extremely difficult task. 

The task is to identify structures on the  raw CMB maps that deviate from the expected pattern in the standard cosmological model \cite{4}. It is necessary to develop and implement an approach that allows to identify various heterogeneous structures from space maps.  We called them foreground outliers because these structures can be caused by various astrophysical phenomena, such as galaxy clusters, supernovae, or other celestial objects, whose intensity exceeds that of the CMB. Examples of such  objects include galaxies or galaxy clusters, bright stars, supernovae,  dust contaminations, and other celestial objects. The study of  foreground outliers on CMB maps is challenging due to their volume and complexity, requiring the development of methods and technologies to automate and enhance their detection and subsequent analysis. This makes it difficult to find  them all manually.

To address this issue, it is proposed to use machine learning methods, particularly neural networks and clustering algorithms. Neural networks are intended to extract features from CMB maps. Feature extraction involves identifying and extracting important features or patterns in a dataset relevant to the problem. In this case, neural networks will be trained to identify patterns on  CMB maps indicating the presence of foreground outliers. Clustering algorithms will be used to group similar structures in the feature space created by the neural network. These methods can enhance the efficiency and accuracy of  foreground outliers detection on CMB maps.

In machine learning, foreground outliers detection can be classified as the task of detecting anomalous images in a sample consisting of a set of images that belong to some general population. Anomaly detection in arbitrary data can be classified as a binary or multi-cluster clustering task, depending on the purpose (detection or classification). Anomalies are data that do not conform to the established concept of normal behavior. It is important to distinguish anomalies from noise in the data. So, we define foreground outliers as statistically rare patterns on the map regardless of their origin. This means that as a result, among foreground outliers we can observe both point-like objects of known or unknown astrophysical origin, and might be CMB regions that do not correspond to the Lambda-CDM model.

Anomaly detection is characterized by unique features that do not allow typical clustering approaches to be applied without modification. These features depend on the task statement and the subject area, while the data structure also has an impact. An overview of the main methods for anomaly detection is provided in \cite{6}.

The main features of the anomaly detection task include:

\begin{enumerate} 
  \item Class imbalance, where anomalous objects constitute a small fraction of the total data (usually less than 1\%).
  \item The potential absence of anomalies in the training sample, while their appearance in real data necessitates effective model detection.
  \item Difficulties in defining a universal measure of similarity between data samples.
  \item The challenge of distinguishing noise from anomalies in data, leading to poor anomaly detection quality in noisy datasets.
\end{enumerate}

In this paper, the problem of detecting anomalies in images is considered. There are several methods for solving this problem. The work \cite{7} provides an overview of the main anomaly detection methods. Here are some of them:

\begin{enumerate} 
  \item Statistical methods. One approach involves constructing a sample distribution function and identifying points that fall outside this distribution.
  \item Feature extraction-based methods. Machine learning algorithms, such as neural networks, can identify patterns in data \cite{8}, \cite{9}. By analyzing these patterns, indicators of anomalies can be found. This approach is particularly effective for high-dimensional data.
  \item Reconstruction-based methods. Autoencoders can be used to reconstruct images, and reconstruction errors can indicate anomalies.
  \item Clustering-based methods. Clustering algorithms can group similar data, with any image that does not fit the clusters being considered an anomaly.
\end{enumerate}

The choice of method depends on the specific task and characteristics of the data, so first you need to analyze the available data.

\section{Data Analysis}

\begin{figure}
\centering
\includegraphics[width=1\linewidth]{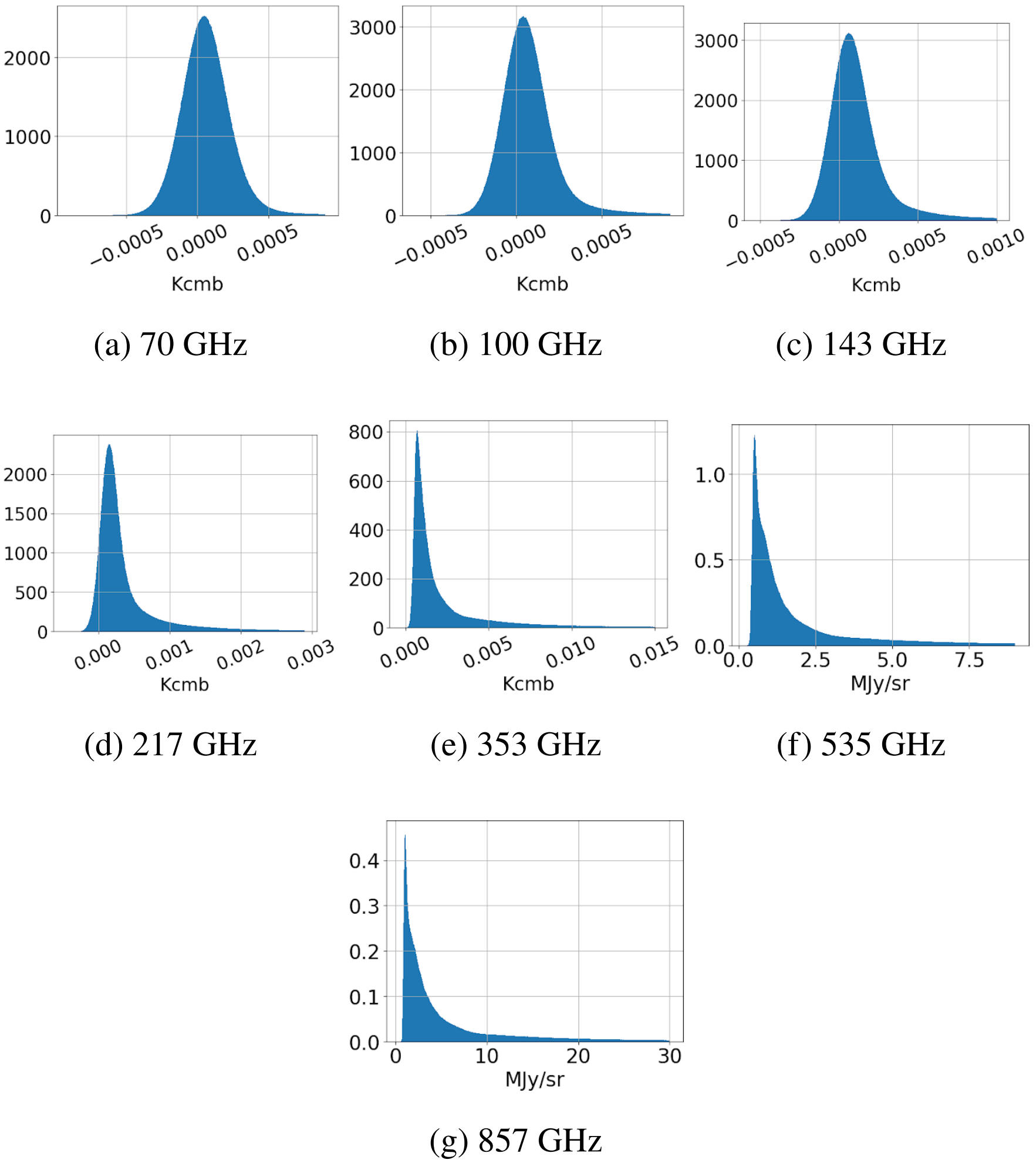}
\caption{Histograms of the intensity of the source data}
\label{fig:1}
\end{figure}

The initial data for the analysis of the relic spectrum were taken from the website of the Planck mission \cite{10}. The total data sample size is about 350 million measurements. All sky maps are presented in the HEALPix format (Hierarchical Equal Area isoLatitude Pixelization) \cite{11}, with $N_{side}$ 1024 or 2048 \cite{12}, in galactic coordinates and with nested order. The data is presented as a set of images, where each image corresponds to a map of the relic radiation at a certain frequency. The signal is given in units of $K_{CMB}$ for 30–353 GHz or MJy/sr ($1 \text{Jy} = 10^{-26}\text{W}/(\text{m}^2 \cdot \text{Hz})$) for constant energy distribution $\nu F_\nu$ for 545 and 857 GHz. The data includes maps of the relic radiation at frequencies 30, 44, 70, 100, 143, 217, 353, 545, and 857 GHz \cite{10}. Maps at frequencies from 70 GHz to 857 GHz were selected for analysis, as they have $N_{side}=2048$ and a relatively low level of instrumental noise. Figure \ref{fig:1} shows histograms of the intensity distribution of the relic radiation at the analyzed frequencies. To gain a better understanding of the data distribution, the long tails of the distributions have been removed from the histograms.




\begin{figure}
\centering
\includegraphics[width=1\linewidth]{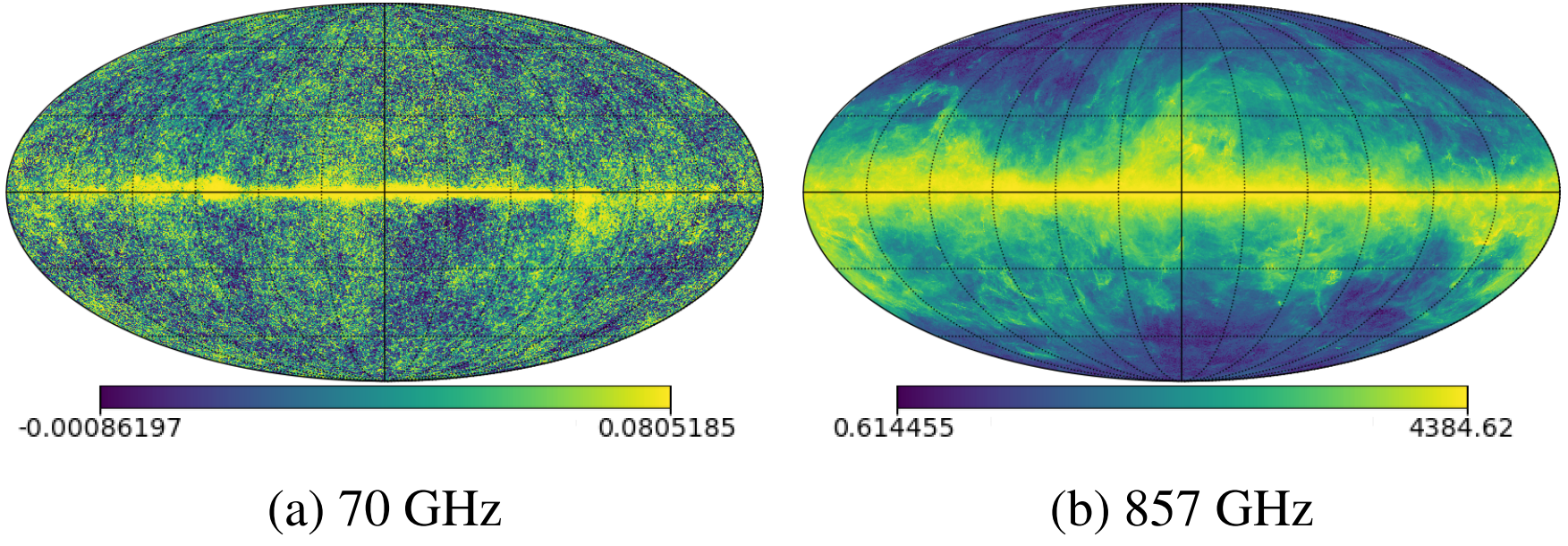}
\caption{Visualization of the source data}
\label{fig:2}
\end{figure}

For visualization and processing of spherical pixel data, the healpy library of the Python language was used, which allows efficient processing of data in the HEALPix format \cite{14}. Figure \ref{fig:2} shows a map of the relic radiation at frequencies of 70 and 857 GHz, obtained using healpy.


Before the analysis, the data is preprocessed and transformed. It is important to correct for galactic foreground radiation, which can distort the signal of the relic radiation. For this reason, the central band of 10 degrees is removed from the maps of the relic radiation, since the Milky Way galaxy can lead to background contamination. The initial data also includes a high level of instrumental noise \cite{15} associated with the instability of the HEMT amplifier (High Electron Mobility Transistor) and other effects. To reduce it, Gaussian smoothing is applied to the data. This method involves convolving the data with a Gaussian kernel, which smooths out small fluctuations. The kernel size is set to 0.15 degrees to suppress noise while maintaining the signal structure. The initial distributions are then normalized:

\begin{equation} \label{eq:1}
\hat{\xi} = \frac{\xi - \mu}{\sigma}
\end{equation}

\begin{equation} \label{eq:2}
\hat{\xi} = \frac{log \xi - \mu}{\sigma}
\end{equation}

\begin{equation} \label{eq:3}
\hat{\xi} = 4 \frac{log \xi - z_{0.5}}{z_{0.975} - z_{0.025}},
\end{equation}
where $\xi$ is a random variable from the initial distribution, $\mu$ is the expectation estimate, $\sigma$ is the variance estimate, and $z_p$ is the quantile of the p level.

\begin{figure}
\centering
\includegraphics[width=1\linewidth]{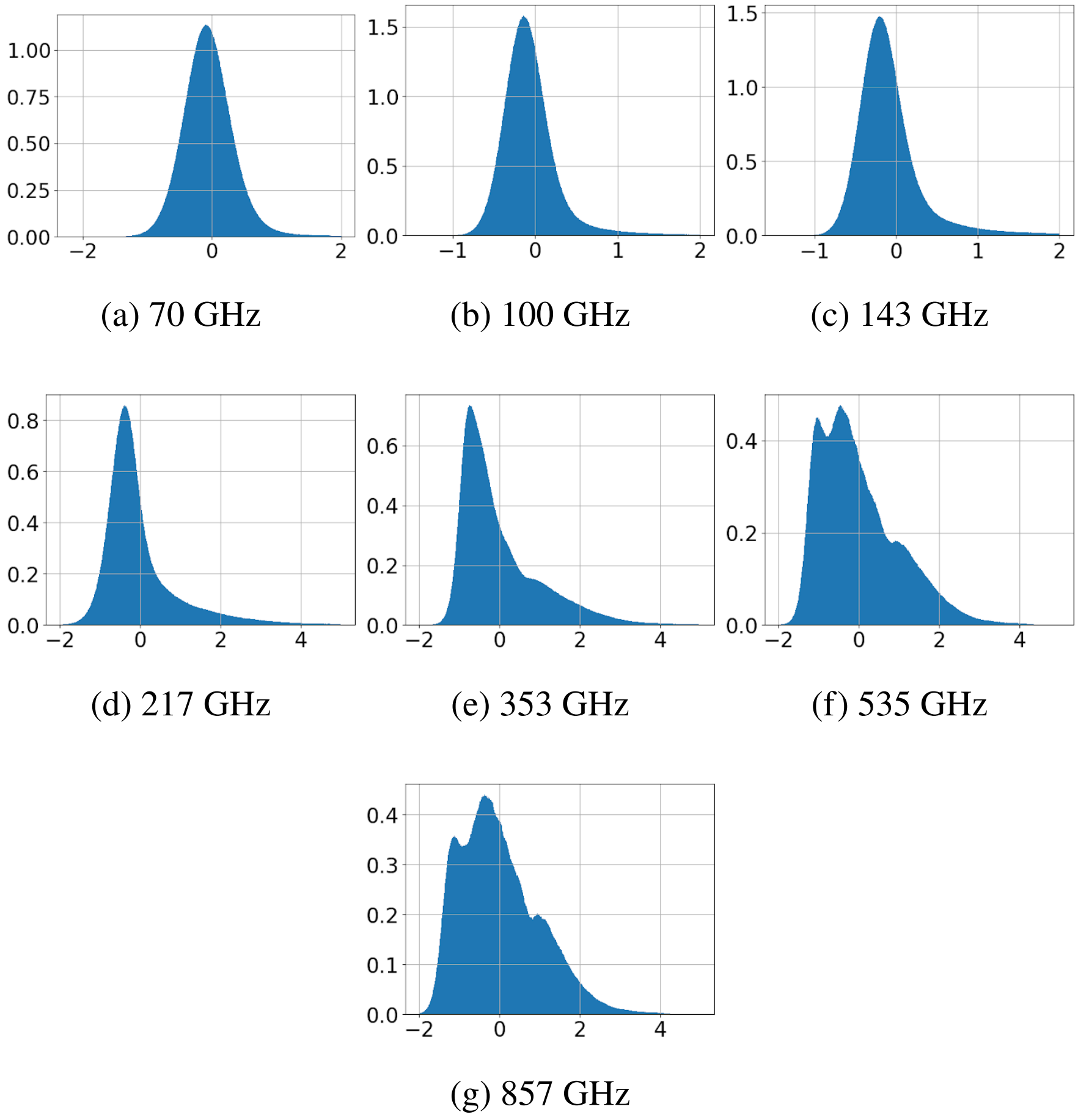}
\caption{Histograms of the intensity of the transformed data}
\label{fig:3}
\end{figure}

The distributions at frequencies of 70, 100, and 143 GHz are close to normal (see Fig. \ref{fig:1}a-\ref{fig:1}c) and scaled according to the formula (\ref{eq:1}). The distribution at a frequency of 217 GHz is close to the log-normal distribution (see Fig. \ref{fig:1}d), therefore the formula (\ref{eq:2}) is used. The distributions at frequencies 353, 545, and 857 GHz are distributions with heavy tails (see Fig. \ref{fig:1}e-\ref{fig:1}g) and are scaled according to the formula (\ref{eq:3}). After the transformation, the histograms of the distribution of the intensity of the relic radiation took the form shown in Fig. \ref{fig:3}. For all frequencies, the data is normalized so that 95\% of all data lies in the range $[-2;2]$.




To create a training sample

\begin{equation} \label{eq:4}
\left\{X_i\right\}_{i=1}^K
\end{equation}

random sections of the map with a size of 2 angular degrees are projected onto a plane, resulting in an image with a resolution of $128\times 128$ pixels. For each point defined by latitude and longitude, there are 7 images that will be analyzed for  foreground outliers. The procedure is repeated K=1536000 times to create a comprehensive training sample. To regularize \cite{16}, noise and zeroing of random pixels are applied to the training sample:

\begin{equation} \label{eq:5}
\left\{s\left(X_i\right)\right\}_{i=1}^K,
\end{equation}
where $s:\mathbb{R}^{7 \times 128 \times 128} \rightarrow \mathbb{R}^{7 \times 128 \times 128}$, $s(x) = \tau\left(x+n\right)$, $n \sim \mathcal{N}(0,0.02)$ and $\tau$ is a function that zeroes random 50\% of the signal for each input value.

\begin{figure}
\centering
\includegraphics[width=1\linewidth]{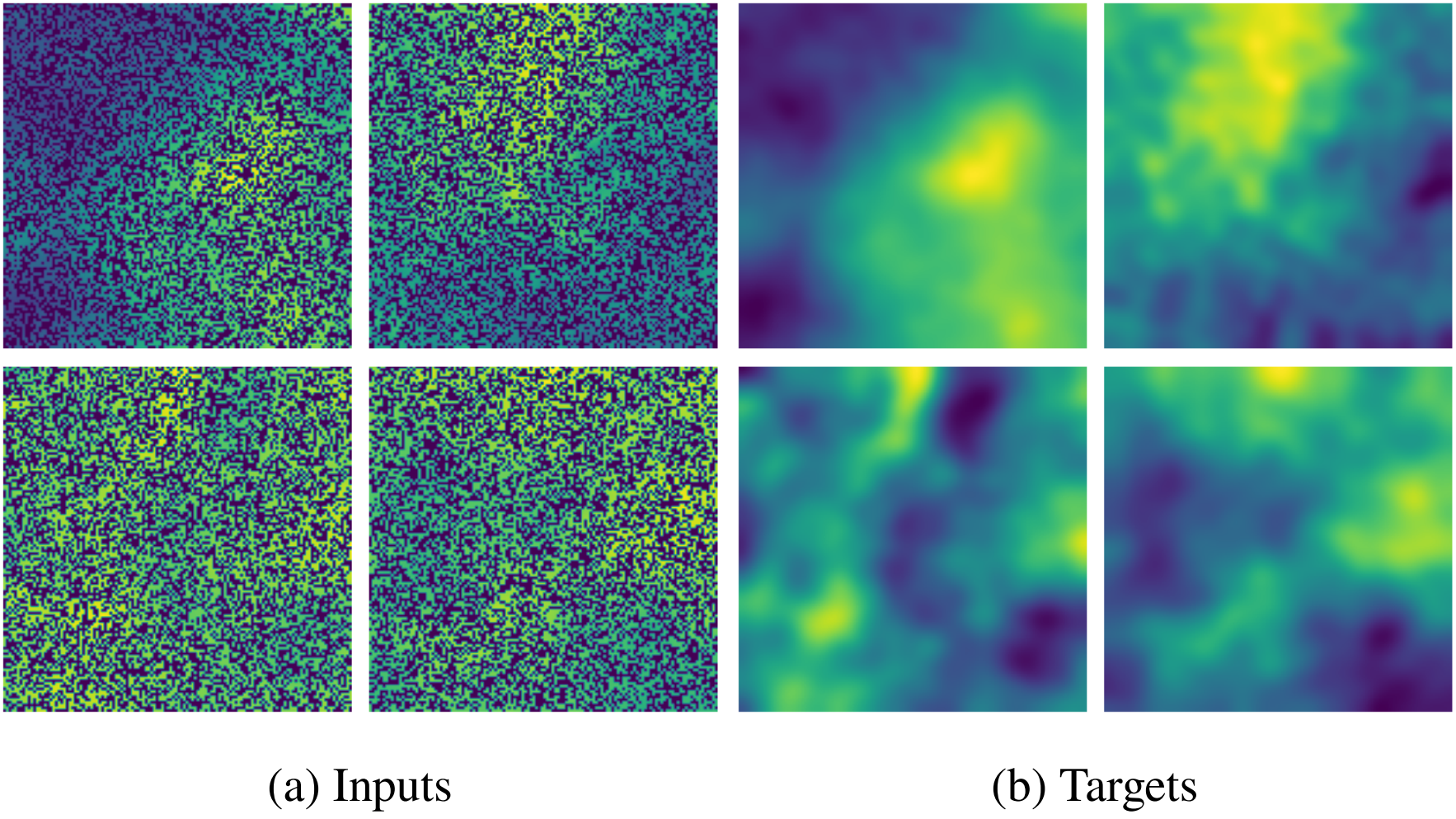}
\caption{Example of preprocessed data}
\label{fig:4}
\end{figure}

It is important to note that the target data for neural network training will be $\left\{X_i\right\}_{i=1}^K$, while the input features will be $\left\{s\left(X_i\right)\right\}_{i=1}^K$. Figure \ref{fig:4}a shows examples of input features (noisy images). Corresponding examples of target values (smoothed images) are shown in Fig. \ref{fig:4}b.


\section{Machine Learning System Architecture}

An auto-encoder \cite{16} is used to extract features from fragments of CMB maps. The vectors from the feature space are then fed into another model responsible for  foreground outlier detection. One of three  foreground outlier detection models is proposed: a statistical model, a clustering model, or an error reconstruction model.

An auto-encoder consists of two parts: an encoder and a decoder. The encoder is a mapping of

\begin{equation} \label{eq:6}
f_{\theta_1}: \mathbb{R}^{7 \times 128 \times 128} \rightarrow \mathbb{R}^N.
\end{equation}

The purpose of the encoder is to transform the input image $X_i$ into a feature vector $f_{\theta_1}\left(X_i\right) \in \mathbb{N}$with a given set of parameters $\theta_1$.  The decoder is a mapping of

\begin{equation} \label{eq:7}
g_{\theta_2}: \mathbb{R}^N \rightarrow \mathbb{R}^{7 \times 128 \times 128}.
\end{equation}

The purpose of the decoder is to reconstruct the feature vector $f_{\theta_1}\left(X_i\right) \in \mathbb{R}^N$ back to the original image $X_i \approx g_{\theta_2} \left( f_{\theta_1}\left(X_i\right) \right) \in \mathbb{R}^{7 \times 128 \times 128}$ with minimal error using a set of parameters $\theta_2$. This way, the neural network learns to create compressed vector representations of images and then reconstruct them with minimal loss. To obtain a higher-quality latent space containing more information about the original images, it is useful to employ denoising auto-encoders. In this approach, a noisy image $s(X_i)$ is input, which serves as a form of regularization and prevents the neural network from overfitting to certain features while ignoring others. Typically, normal or uniform noise with low variance is applied. In some cases, certain input values are reset, equivalent to using dropout \cite{18} in the input layer.

The VGG19 model was used as the base model, enabling the network to extract features from images at different scales \cite{20}. The encoder consists of 5 main blocks, each containing two or four conv3 blocks. Each block starts with a convolution layer with a 3×3 kernel, a stride of 1, and a padding of 1. Each convolution layer is followed by a batch normalization layer and a LeakyReLU activation function. Each main block is followed by a max-pooling block that reduces the dimensions by a factor of 2. This block is similar to the conv3 block but has a 2×2 convolution kernel, a stride of 2, and no padding. Dropout layers \cite{18} are also included between the main blocks. These modifications allow for the correct processing of 7×128×128 tensors instead of the original 3×224×224 tensors, maintaining an acceptable batch size for normalization layers. Another change is in the last fully connected layer, which has N outputs corresponding to the feature vector dimension, with the activation function replaced by Tanh. The decoder has a similar architecture to the encoder but with the blocks arranged in reverse order. Instead of convolutional layers, fractional step convolution layers (or deconvolution layers) are used to increase the dimensions of objects \cite{22}. The model was built using the PyTorch machine learning framework.

The auto-encoder was trained using the Adam method \cite{24} with a mini-batch size of 32. Xavier initialization \cite{25} with a normal distribution was used for the initial weight setup. For normalization layers, a normal distribution with a mean of 1 and low variance was used for initialization. All free coefficients were initialized to zero. Experiments with various models showed that a linear combination of the mean squared error and Kullback-Leibler divergence is suitable as a loss function:

\begin{equation} \label{eq:8}
    L(y, \hat{y}) = 5L_2(y, \hat{y}) + D_{KL}(y, \hat{y}),
\end{equation}

\begin{equation} \label{eq:9}
    L_2(y, \hat{y}) = \frac 1 2 \sum_{k=1}^{7} \sum_{j,i=1}^{128} \left( y_{kji} - \hat{y}_{kji} \right)^2,
\end{equation}

\begin{equation*}
D_{KL}(y, \hat{y}) = \sum_{k=1}^{7} \sum_{j,i=1}^{128} 
\left( y_{kji} log \frac {y_{kji}}{\hat{y}_{kji}} \right),
\end{equation*}
where $y \in \mathbb{R}^{7 \times 128 \times 128}$  is the target image, and $\hat{y} \in \mathbb{R}^{7 \times 128 \times 128}$ is the output of the auto-encoder. Training with only $L_2$ as a loss function led to smaller deviations in image areas with the highest and lowest radiation intensity but resulted in a distribution significantly different from expected. Using the only $D_{KL}$ loss solved the overall distribution deviation problem but led to significant deviations in some image areas, primarily those with the highest and lowest radiation intensity. Increasing the $L_2$ weight in the loss function caused greater granularity in the restored images, while increasing the $D_{KL}$ weight resulted in insufficient image contrast.

Thus, the training process aims to minimize:
$$
\sum_{i=1}^K L\left( X_i, 
g_{\hat{\theta}_2} \left( 
f_{\hat{\theta}_1} \left( 
s \left(  \hat{X}_i \right) \right) \right) \right) \rightarrow \min_{\theta_1, \theta_2}
$$
where the input value is taken from (\ref{eq:5}), and the target value corresponds to the input from the sample (\ref{eq:4}). The learning rate parameter was halved whenever the error plateaued. Cyclic parameter changes between the boundaries of a triangular linear window (the triangle method) described in \cite{26} showed slightly worse results.

During the Leslie Smith procedure \cite{27} for determining the optimal learning rate, two intervals were found: $[0,01;0,2]$ and $[5\cdot 10^{-5};5\cdot 10^{-4}]$. In the first interval, the network converged to a trivial solution, while a stable non-trivial solution was found in the second interval.

\begin{figure}
\centering
\includegraphics[width=1\linewidth]{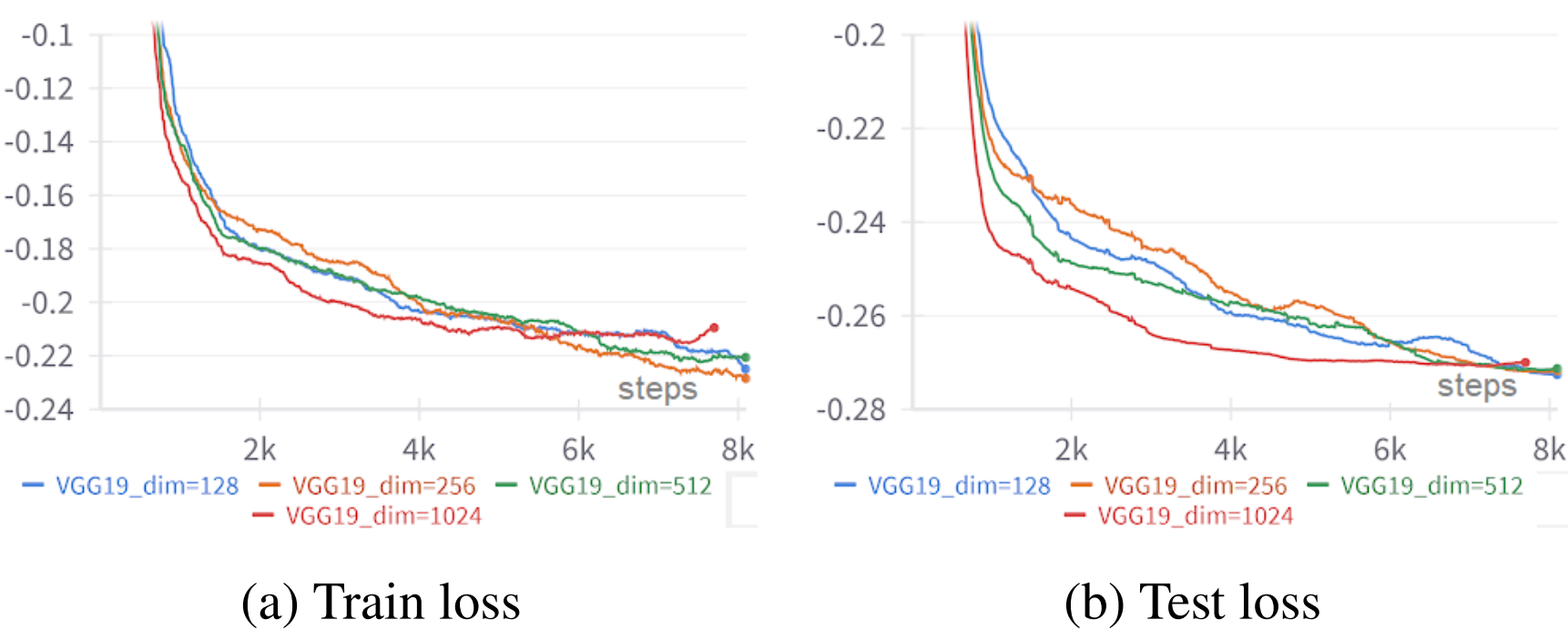}
\caption{The training losses}
\label{fig:8}
\end{figure}

Figure \ref{fig:8} shows the network training process. Figure \ref{fig:8}a shows the error change in the training sample, and Figure \ref{fig:8}b shows the error change in the test sample. During training, it was found that increasing the parameter $N > 128$ does not significantly improve the error functionality. Thus, a network with $N=128$ will be used to optimize the number of further calculations. Let the parameters of the pretrained encoder network (\ref{eq:6}) be denoted as $\hat{\theta}_1$, and the decoder parameters (\ref{eq:7}) as $\hat{\theta}_2$.


Starting from a certain value of parameter N, further increases in dimensionality do not affect model quality. However, high dimensionality of feature vectors is unjustified and can significantly slow down  foreground outlier detection models. Training multiple auto-encoder models to determine the optimal N value for each task is also too resource-intensive. Therefore, principal component analysis \cite{28} can be used to reduce the feature space dimensionality to improve  foreground outlier detection model performance. However, if the number of dimensions is too low, this method can negatively impact  foreground outlier detection accuracy.

The paper proposes using principal component analysis based on SVD decomposition, but for large feature spaces, randomized truncated SVD decomposition \cite{29} is necessary.

If the sample $\left\{f_{\hat{\theta}_1} \left( \hat{X}_i \right) \right\}_{i=1}^P$ is centered, i.e., $\mathbb{E} \left[f_{\hat{\theta}_1} \left( \hat{X}_i \right) \right] = 0$, then by reducing the dimensionality to the first $l \leq N$ principal components with eigenvalues $\lambda_1 \geq \ldots \geq \lambda_l \geq \ldots \geq \lambda_N \geq 0$, the ratio $\Lambda = \frac{\lambda_1 + \ldots + \lambda_l} {\lambda_1 + \ldots + \lambda_N}$ of explained variance to sample variance can be used to find the optimal dimensionality of the feature space.

If the sample is not centered, this can always be achieved by defining a new $\widetilde{f}_{\hat{\theta}_1} \left( \hat{X}_i \right) = f_{\hat{\theta}_1} \left( \hat{X}_i \right) + C$ where $C = - \frac 1 P \sum_{i=1}^P f_{\hat{\theta}_1} \left( \hat{X}_i \right)$.

\begin{figure}
\centering
\includegraphics[width=.8\linewidth]{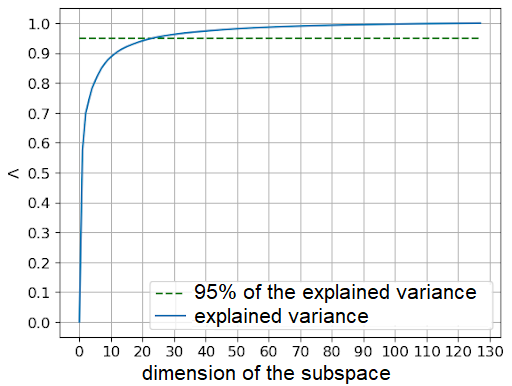}
\caption{The explained variance}
\label{fig:7}
\end{figure}

The proportion of explained variance in the centered sample $\left\{f_{\hat{\theta}_1} \left( \hat{X}_i \right) \right\}_{i=1}^P$, i.e., $\mathbb{E} \left[f_{\hat{\theta}_1} \left( \hat{X}_i \right) \right] = 0$, depending on the subspace dimensionality, is shown in Fig. \ref{fig:7}. The figure shows that approximately 25 dimensions are sufficient to explain 95\% of the total sample variance. Using SVD decomposition allows for training an auto-encoder with a sufficiently large feature space dimension once, and then reducing the dimensionality on the required sample with minimal loss of feature quality.

\section{Foreground Outliers Detection Models}

After training the neural network, a new dataset was compiled:

\begin{equation} \label{eq:10}
\left\{\hat{X}_i\right\}_{i=1}^K.
\end{equation}

This dataset was used to identify  foreground outliers. The sample construction followed the training sample algorithm (\ref{eq:4}), but for different random sections of the map. The sampling procedure was repeated K=10240 times, with no image noise applied.

At this stage, we have a trained autoencoder model. Therefore, alongside the sample (\ref{eq:10}), the corresponding samples can be used to detect  foreground outliers:

\begin{equation} \label{eq:11}
\left\{f_{\hat{\theta}_1}\left(\hat{X}_i\right)\right\}_{i=1}^K,
\end{equation}

\begin{equation} \label{eq:12}
\left\{g_{\hat{\theta}_2}\left(f_{\hat{\theta}_1}\left(\hat{X}_i\right)\right)\right\}_{i=1}^K.
\end{equation}

These samples were obtained using the pre-trained encoder (\ref{eq:6}) and decoder (\ref{eq:7}) based on the VGG19 network, trained on sample (\ref{eq:4}).

To detect  foreground outliers, we propose three models: a statistical model, a DBSCAN clustering model, and a model analyzing autoencoder reconstruction errors.

The statistical model analyzes the data distributions in sample (\ref{eq:11}) and identifies points that deviate from this distribution. This model is used to detect  foreground outliers in feature vectors created by the autoencoder. The algorithm for the statistical model is as follows:

\begin{enumerate} 
  \item For the given sample (\ref{eq:11}) of feature vectors
  $\left\{f_{\hat{\theta}_1}\left(\hat{X}_i\right) = \left( f_{\hat{\theta}_1}\left(\hat{X}_i\right)^1, \dots, f_{\hat{\theta}_1}\left(\hat{X}_i\right)^N \right)^T \right\}_{i=1}^K$, for each projection $\left\{f_{\hat{\theta}_1}\left(\hat{X}_i\right)^j \right\}_{i=1}^K$ onto the j-th element of the feature space basis, an estimate of the probability density function is constructed.
  \item For each $j=1\dots N$, the q-th confidence interval $\left[z_{\frac{1-q}{2}};z_{\frac{1+q}{2}}\right]$ is constructed, and sample elements $\left\{f_{\hat{\theta}_1}\left(\hat{X}_i\right)^j \right\}_{i=1}^K$ that do not lie within the interval are identified.
  \item For each $i=1 \dots P$, it is determined whether element $\hat{X}_i$ is anomalous  (in the sense of machine learning). If $k \in \mathbb{N}$ or more projections $f_{\hat{\theta}_1}\left(\hat{X}_i\right)^j$ of the element 
 $f_{\hat{\theta}_1}\left(\hat{X}_i\right)$ in sample (\ref{eq:11}) do not belong to the corresponding intervals, then the i-th element is considered anomalous.
\end{enumerate}

\begin{figure}
\centering
\includegraphics[width=1\linewidth]{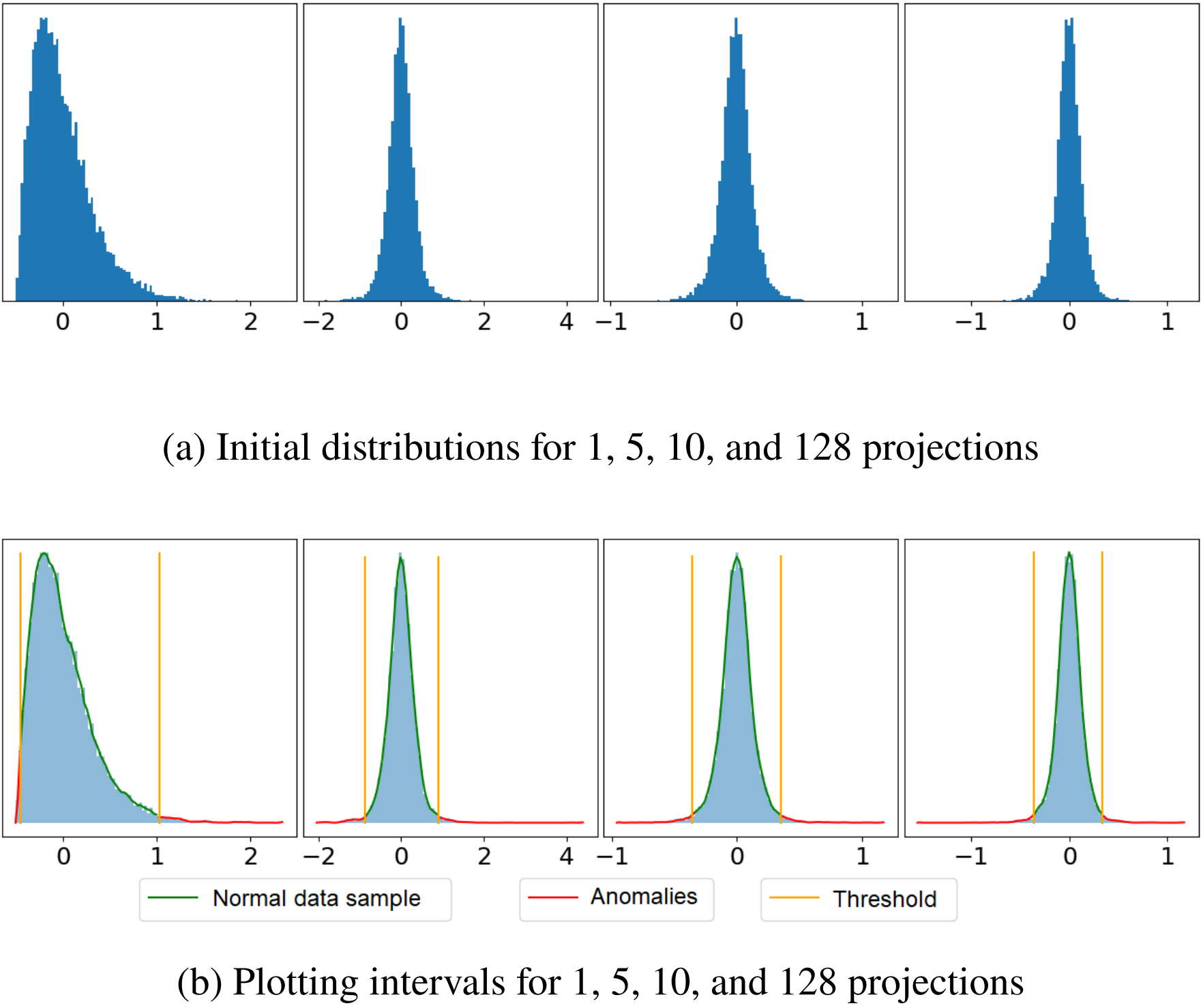}
\caption{Algorithm illustration}
\label{fig:9}
\end{figure}

Figure \ref{fig:9} illustrates the algorithm for sample (\ref{eq:11}) with a 95\% confidence interval. Figure \ref{fig:9}a shows the distribution across some feature space projections for sample (\ref{eq:11}), corresponding to the first step of the algorithm. Figure \ref{fig:9}b shows the construction of confidence intervals and the separation of the distribution into normal and abnormal parts for each projection, corresponding to the second step of the algorithm.

$q \in \left[ 0.5;1\right]$ and $k \in \left[ 1;N \right]$ are model parameters. The algorithm has an asymptotic complexity of $O(KN)$.


The second approach to  foreground outliers detection is using clustering algorithms that allow for abnormal data samples. We hypothesize that normal objects form one dense cluster, while abnormals are sparsely distributed on the periphery. Models such as DBSCAN (Density-Based Spatial Clustering of Applications with Noise) \cite{30}, Isolation Forest (IF) \cite{31}, or Local Outlier Factor (LOF) \cite{32} can be used.

DBSCAN, a density-based method, was chosen as the clustering algorithm. Clusters are high-density regions in space, separated by low-density regions. The DBSCAN clustering algorithm can be described with the following definitions:

The epsilon neighborhood of an image $\hat{X}_i$ from sample (\ref{eq:10}) is the set $U_{\varepsilon}\left(\hat{X}_i\right) = \left\{ \hat{X}_j: j=1\dots P, \lVert f_{\hat{\theta}_1}\left(\hat{X}_i \right) - f_{\hat{\theta}_1}\left(\hat{X}_j \right)\rVert < \varepsilon \right\}$. The image $\hat{X}_i$ is considered to lie in its neighborhood. An image $\hat{X}_i$ from sample (\ref{eq:10}) is a core point if $U_{\varepsilon}\left(\hat{X}_i\right) \geq m, m \in \mathbb{N} $. An image $\hat{X}_j$ is directly density-reachable from $\hat{X}_i$ if $\hat{X}_j \in U_{\varepsilon}\left(\hat{X}_i\right)$ and $\hat{X}_i$ is a core point. An image $\hat{X}_j$ is density-reachable from $\hat{X}_i$ if there exists a chain of images $I_0, I_1, \dots, I_n$ from sample (\ref{eq:10}) such that $I_0 = \hat{X}_i, I_n = \hat{X}_j$, and $\forall k = 1, \dots , n$, the image $I_k$is directly density-reachable from $I_{k-1}$. Clusters $S_1,S_2 \subset \left\{ \hat{X}_i \right\}_{i=1}^K$ are density-reachable if all core points from cluster $S_1$ are density-reachable from the core points of cluster $S_2$ and vice versa. An image $\hat{X}_i$ is (directly) density-reachable from cluster $S \subset \left\{ \hat{X}_i \right\}_{i=1}^K$ if it is (directly) density-reachable from a core point in $S$.

The DBSCAN clustering algorithm can be summarized as follows:
\begin{enumerate} 
  \item For each core image $I \in \left\{ \hat{X}_i \right\}_{i=1}^K$, place it in cluster $S \subset \left\{ \hat{X}_i \right\}_{i=1}^K$, along with all images directly density-reachable from $I$.
  \item Repeat steps 3 and 4 until cluster stabilization.
  \item If there are density-reachable clusters $S_1,S_2 \subset \left\{ \hat{X}_i \right\}_{i=1}^K$, combine them.
  \item For all remaining clusters $S_i$, include all directly density-reachable images not yet included, if such images exist.
\end{enumerate}

\begin{figure}
\centering
\includegraphics[width=1\linewidth]{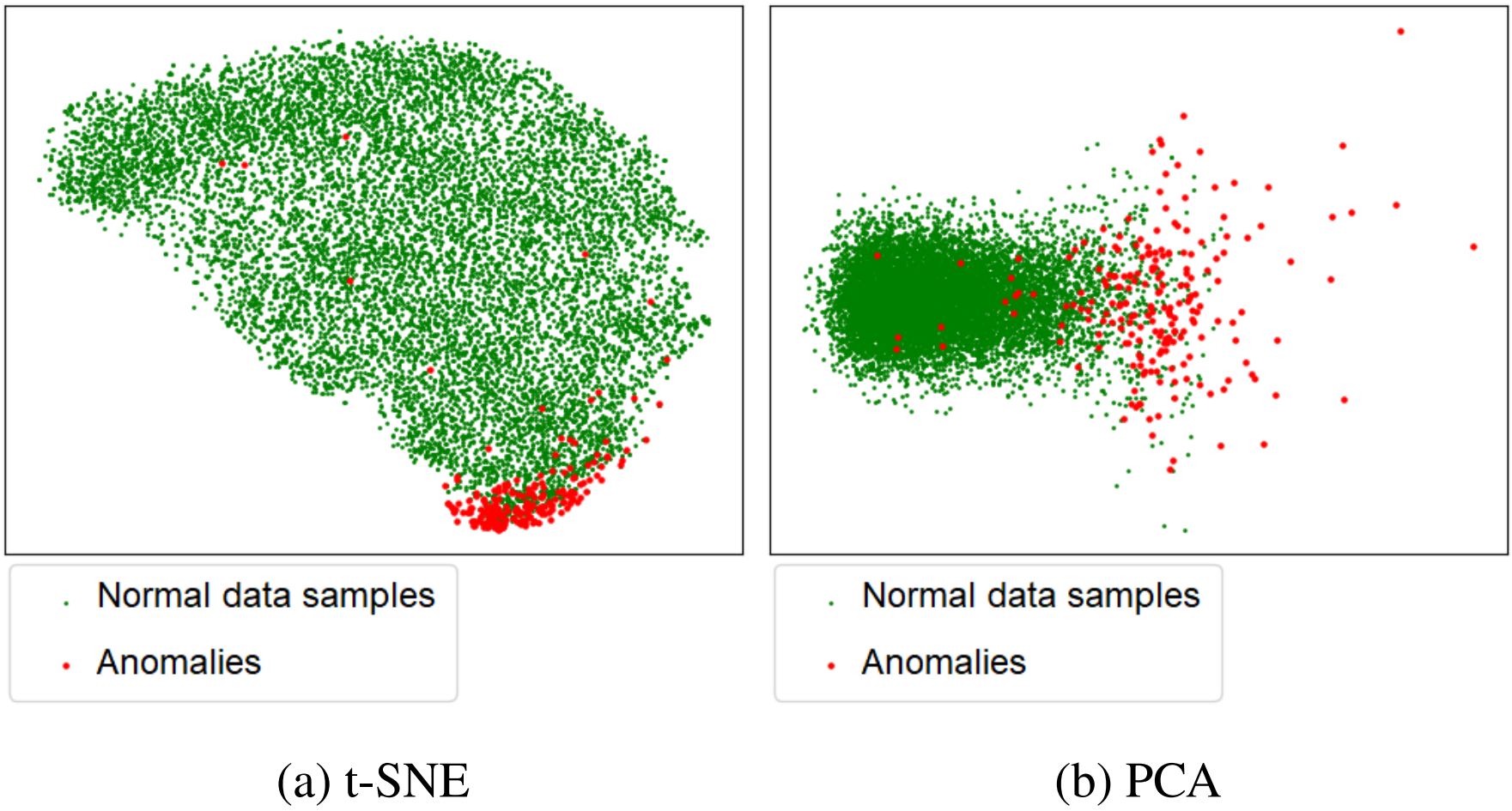}
\caption{Visualization of foreground outliers in low-dimensional space}
\label{fig:10}
\end{figure}

The algorithm parameters $\varepsilon \in \mathbb{R}$ and $m \in \mathbb{N}$ define the neighborhood radius and minimum points in this neighborhood. Parameters were optimized using GridSearchCV \cite{34} to ensure a non-empty set of  abnormal objects and validate the object distribution hypothesis. This procedure classifies all objects into normal and abnormal. Figure \ref{fig:10} shows the division of sample (\ref{eq:10}) into normal and abnormal images in low-dimensional space, using t-SNE (Fig. \ref{fig:10}a) \cite{35} and Sparse PCA (Fig. \ref{fig:10}b) \cite{36}.


The figures show that  foreground outliers also form a cluster. This may result from dimensionality reduction during data visualization. Anomalous images  in sense of machine learning with similar visual features may cluster in feature space due to similar vector representations. This effect, studied in \cite{37}, \cite{38}, \cite{39}, causes anomalies to form low-density clusters, which DBSCAN might not detect.

The error reconstruction model assumes the autoencoder was primarily trained on normal data, enabling accurate reconstruction of normal images per the error function (\ref{eq:8}). For abnormal data, the reconstruction error (\ref{eq:8}) is expected to be higher due to their scarcity in the training set. The training process prioritized minimizing overall error by effectively compressing normal data, resulting in less effective compression of  foreground outliers. Besides the encoder output (\ref{eq:11}), the decoder output (\ref{eq:12}) can be used. For each image $\hat{X}i$ in sample (\ref{eq:10}), the reconstructed image $g_{\hat{\theta}_2} \left( f_{\hat{\theta}_1} \left( \hat{X}_i \right) \right)$ can be compared to the original image. Any known error function can be used, but this work uses functional (\ref{eq:9}), which was employed during autoencoder training for its symmetry, accuracy, and non-negative definiteness.

\begin{figure}
\centering
\includegraphics[width=1\linewidth]{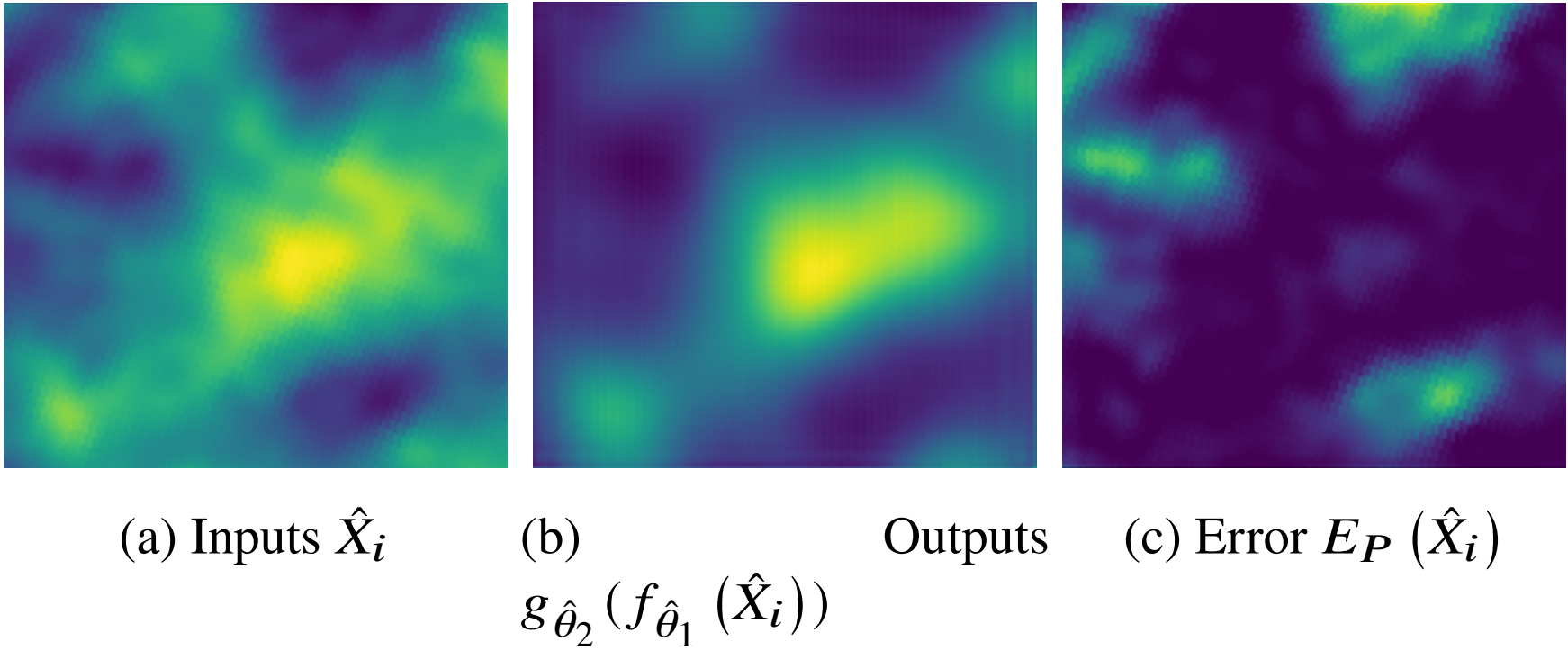}
\caption{An example of calculating the reconstruction error}
\label{fig:11}
\end{figure}

An example of calculating reconstruction error for one image is shown in Fig. \ref{fig:11}a, with the input image. Fig. \ref{fig:11}b shows the image obtained by compressing to low-dimensional space and restoring with the autoencoder.

A key advantage of the error reconstruction model is its ability to detect  anomalous images and accurately identify pixels deviating most from the expected distribution. Fig. \ref{fig:11}c shows the pixel-wise error 
$$
E_P \left( \hat{X}_i \right) = vec^{-1} \left( \left[ vec \left( \hat{X}_i \right) - vec \left( g_{\hat{\theta}_2} \left( f_{\hat{\theta}_1} \left( \hat{X}_i \right) \right) \right) \right]^2 \right)
$$
between the input and reconstructed images, where $vec$ is the image-to-vector transformation, and squaring is coordinate-wise The reconstruction error is the sum of $E_P \left( \hat{X}_i \right)$ errors:
$$E_r \left( \hat{X}_i \right) = \frac{1}{128^2} \sum_{j,k=1}^{128} \left[ E_P \left( \hat{X}_i \right) \right]_{jk}.$$

This result is calculated for each image from the sample (\ref{eq:10}) and reflects the degree of  abnormality of each image: the smaller the error,  the less abnormal the image. Images with an error exceeding the set threshold are considered  contains foreground outlier.


More formally, with the error reconstruction method, $M \in \mathbb{N}$ of the largest relative order of $\leq$ are selected as sample objects  contained foreground outliers (\ref{eq:10}), where $\hat{X}_i \leq \hat{X}_j \Leftrightarrow E_r \left( \hat{X}_i \right) \leq E_r \left( \hat{X}_j \right)$.

\section{Analysis of Results}

To quantify the degree of consistency of the model results, the Matthews correlation coefficient \cite{40} can be used:

$$C_M = \frac{
(T_P \times T_N) - (F_P \times F_N)
}{
\sqrt{(T_P + F_P) \times (T_P + F_N) \times (T_N + F_P) \times (T_N + F_N)}
},$$
where $T_P$ is the number of true-positive results; $T_N$ is the number of true-negative results; $F_P$ is the number of false-positive results; $F_N$ is the number of false-negative results. Table \ref{tab:1} shows the results of pairwise correlations for each pair of models, with the results of one model taken as the true answers. The Matthews correlation coefficient is symmetric and robust to class sizes.

\begin{table}
	\centering
	\caption{Pairwise Correlation of Models}
	\label{tab:1}
	\begin{tabular}{lccc} 
		\hline
$C_M$ & \text{Statistics}  & \text{Clustering} & \text{Reconstruction} \\
		\hline
\text{Statistics}   & 1.0     & 0.975  & 0.955  \\
\text{Clustering}    & 0.975   & 1.0    & 0.968  \\
\text{Reconstruction}    & 0.955   & 0.968  & 1.0    \\
		\hline
	\end{tabular}
\end{table}

To assess the quality of the models, a test sample was compiled, similar to sample (\ref{eq:10}), but consisting of different sections of the Planck mission's CMB maps.

The mask UT78 \cite{41}, which contains information about foreground objects obscuring the CMB such as point sources and other bright regions, is proposed as the target values. This mask is based on data from the Planck mission and includes polarization changes. The mask is a spherical map of zeros and ones with a resolution of $N_{side}=2048$, where zeros denote masked pixels containing foreground objects and ones denote unmasked pixels \cite{42}.

To construct UT78, data obtained using four different methods were utilized: COMMANDER \cite{43}, NILC \cite{44}, SEVEM \cite{45}, and SMICA \cite{46}. Each method employs a different approach to analyzing the intensity and polarization of the CMB and serves to purify the signal from foreground radiation.  It is clear that the UT78 mask does not fully correspond to the foreground outlier regions. On the one hand, the mask contains statistically frequent point-like contaminations. On the other hand, there may be still unknown origin objects that are not contained in the mask. But at the same time, the widest class - point sources of known astrophysical origin from the foregrounds are contained in this mask. Therefore, we compare the results with the mask. The following data are provided for each method \cite{47}:

\begin{enumerate}
    \item Full-scale CMB intensity map with a corresponding confidence mask and effective beam transmission function.
    \item Full-scale CMB polarization map with a corresponding confidence mask and effective beam transmission function.
\end{enumerate}

Thus, the mask of  foreground contamination regions is constructed with more information than was available to the developed model and is consistent with modern understandings of CMB fluctuations \cite{48}. Details of the UT78 construction process are presented in \cite{47}, \cite{48}, and \cite{49}. The latest release is presented in \cite{49}.

\begin{figure}
\centering
\includegraphics[width=1\linewidth]{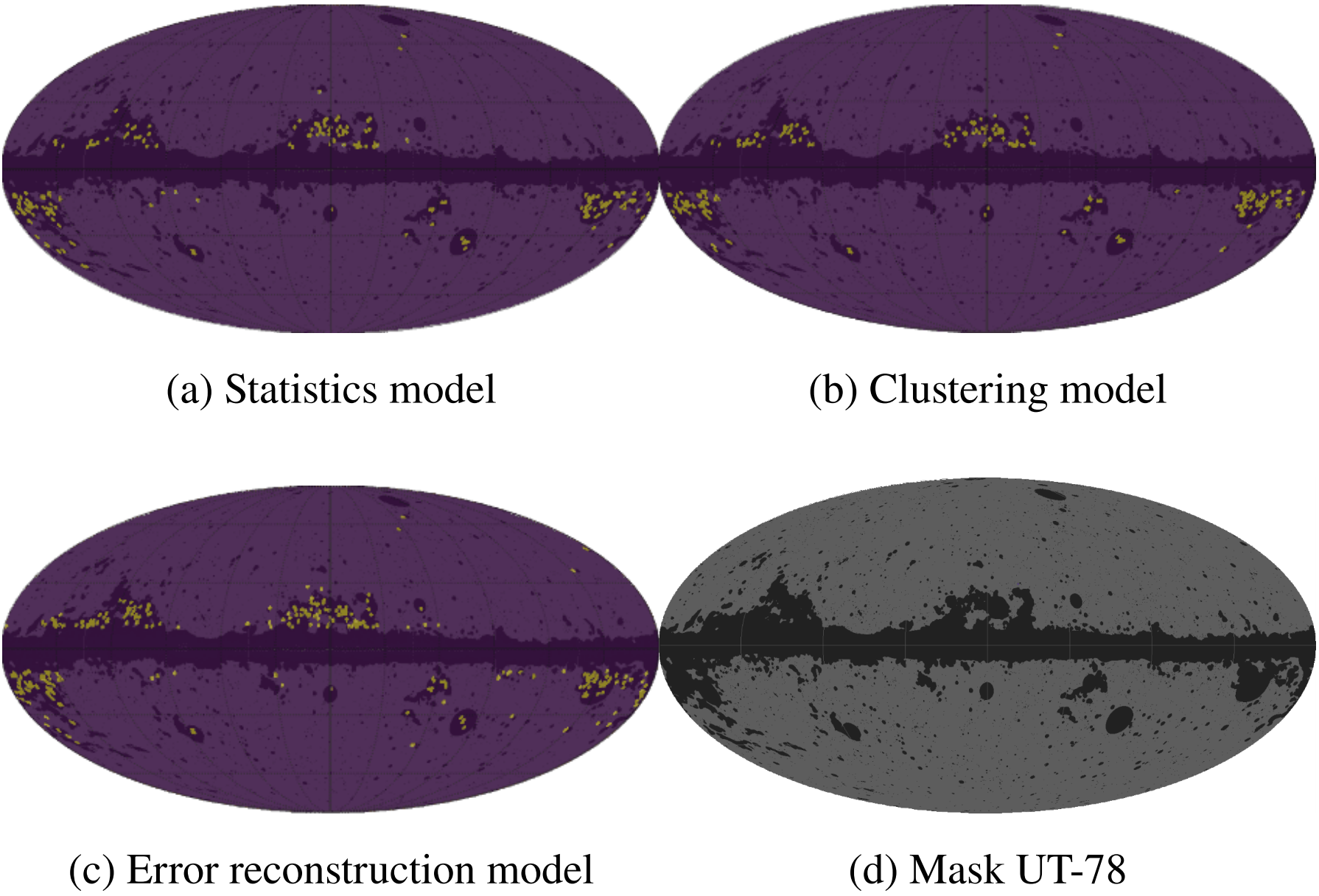}
\caption{Overlay of Found Foreground Outliers on the Mask}
\label{fig:12}
\end{figure}

Figures \ref{fig:12}a, \ref{fig:12}b, and \ref{fig:12}c show the  foreground outliers detected on the Planck mission CMB map \cite{10} using the proposed  foreground outlier detection models with the UT78 mask overlaid. Figure \ref{fig:12}d shows the original UT78 mask. The central band was previously removed from the test sample. It can be seen that most of the detected  foreground outliers correspond to the masked regions in UT78.



The developed models detect  foreground outliers in fixed-size windows, while the mask of  foreground contamination regions provides pixel-by-pixel information. To obtain a correct estimate, it is necessary to create a set of reference data using the mask in a format where each fixed-size window corresponds to an  contamination label. The procedure for building reference data for UT78 was carried out as follows:

\begin{enumerate}
    \item Each window from the test sample is assigned a section of the UT78 mask.
    \item The entire section of the mask is marked as  contaminationed if the proportion of masked pixels exceeds the specified threshold $p \in \left[0;1\right]$. Otherwise, the entire section of the mask is marked as normal.
\end{enumerate}

This procedure allows obtaining target values for the test sample based on the mask of  foreground contamination regions. The threshold $p$ is proposed to be varied. To maintain the class balance in the test sample, the value of $p$ was uniformly distributed. Thus, the proportion of normal and abnormal images is the same at $p=0.5$. In practice, the value of $p$ should not be too large. Ideally, even one masked pixel in the image should indicate that the image contains an  contamination.

However, the mask of  foreground contamination regions was designed such that often, at the analyzed frequencies, foreground objects are masked along with some of their surroundings. This was done because it was more important to prevent radiation from nearby objects from entering the CMB spectrum than to avoid losing a small part of the CMB. This means that for small values of $p$ (e.g., $p < 0.05$), there is a high probability that the area does not actually contain  contamination but does contain masked pixels located close to an  contamination. As a result of testing, the accuracy and completeness of  foreground outliers detection were determined for the threshold $p=0.2$ according to the following formulas:

\begin{equation} \label{eq:14}
    P = \frac{T_P}{T_P+F_P},
\end{equation}

\begin{equation} \label{eq:15}
    A = \frac{T_P}{T_P+F_N},
\end{equation}
where $T_P$ is the number of true positive results, $T_N$ is the number of true negative results, $F_P$ is the number of false positive results, and $F_N$ is the number of false negative results. The metrics (\ref{eq:14}) and (\ref{eq:15}) are not sensitive to class balance.

The results are presented in Table \ref{tab:2}. High accuracy and low completeness mean that most of the detected  foreground outliers are indeed  point-like objects of known astrophysical origin or astrophysical contamination regions, but the models found only a small portion of the masked objects. This may be because some of the masked objects are single dim stars with similar structures, so they are typical of the class of single stars and were not  classified as foreground outliers. This rationale also partially explains why  foreground outliers formed a loose cluster during DBSCAN clustering.

\begin{table}[h!]
\caption{Comparison of Main Model Quality Indicators}
\label{tab:2}
\centering
\begin{tabular}{lccc}
\hline
\textbf{Model} & \textbf{Foreground outliers detected} & \textbf{Precision} & \textbf{Recall} \\
\hline
\textbf{Statistics} & 200 & 0.975 & 0.132 \\
\textbf{Clustering} & 210 & 1.0 & 0.142 \\
\textbf{Reconstruction} & 300 & 0.917 & 0.186 \\
\hline
\end{tabular}
\end{table}

\begin{figure}
  \centering
  \includegraphics[width=1\linewidth]{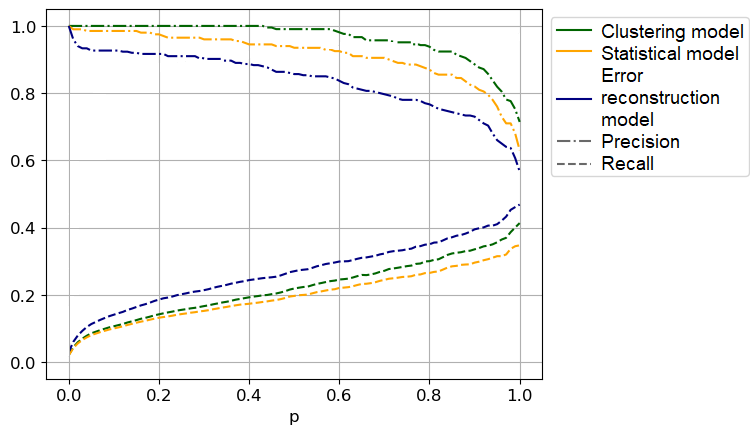}
  \caption{Model Quality Metrics}
  \label{fig:16}
\end{figure}

Figure \ref{fig:16} shows the metrics (\ref{eq:14}) and (\ref{eq:15}) of model quality for all possible values of the parameter $p \in \left[0;1\right]$.

High precision allows the model to be used to search for objects in large volumes of data that stand out from the overall picture and thus potentially interest the researcher. However, it should be noted that the model can only detect some of the  foreground outliers. Low recall indicates that not all areas of the UT78 mask will be marked as  foreground outliers. This is not a significant drawback since the task of constructing a mask for the CMB was not set.

 The developed models demonstrate a high probability of correctly identifying foreground outliers, which predominantly belong to the UT78 mask and possess a known astrophysical nature. This indicates that when the model classifies an image as an anomaly, the likelihood of it containing a foreground outlier is very high. This capability makes the model particularly valuable for analyzing large datasets, where it is necessary to detect objects that deviate from the norm for further investigation.

 The developed ML models identify foreground outliers, primarily part of the UT78 mask, with known astrophysical origins. Unlike many statistical methodologies, these models do not rely on prior information about the astrophysical origin of the sources they aim to detect. This gives the models an advantage in discovering new or unknown objects, whose origin and characteristics have not yet been studied. Such a model could be highly useful for analyzing other maps to detect atypical objects for which no mathematical model exists. Based on the current results, it can be inferred that the detected foreground outliers will mostly form a subset of all the targets, as observed with the UT78 mask.

 However, the model exhibits low recall. This implies that we lack reliable information about the proportion of objects that are true foreground outliers but might be missed by the model. It is possible that a significant portion of the objects within the UT78 mask are statistically rare foreground outliers. In such a scenario, the model's true recall would be low. This feature limits the model's application in tasks that require the identification of all potential foreground outliers until a more accurate estimate of the model's recall is obtained.

One way to evaluate the overall quality of the model, without being tied to the proportion of masked pixels $p$, is ROC-AUC. In our case, not all models are classification models in the usual sense: the DBSCAN model solves a clustering problem that was reduced to classification, and the error reconstruction model solves a ranking problem. Because of this, constructing ROC curves for the models is impossible. However, if the target and predicted class values are swapped, ROC curves can be constructed to illustrate the correspondence between the UT78 mask and each model. In this case, the classification threshold will be the value of $p$.

In other words, we can consider a binary Bayesian classifier that finds the proportion of masked pixels in an image and assigns the input mask to one of two classes according to the following decision rule \cite{50}:

\begin{equation} \label{eq:16}
    l(X) = \frac{P(X|H_0)}{P(X|H_1)} \overset{>}{<} \frac{P(H_1)}{P(H_0)} \Rightarrow X \in \left\{ \begin{array}{l} H_0 \\ H_1 \end{array} \right.
\end{equation}
where $H_0$ is the class of normal images, $H_1$ is the class of abnormal images, and $P(X|H_i)$ is the likelihood function (i=0,1). The likelihood function $P(X|H_i)$ sets the proportion of pixels in the image belonging to class $H_i$. Then the likelihood ratio $l(X)$ can be rewritten as:

$$
l(X) = \frac{P(X|H_0)}{P(X|H_1)} = \left( \frac{\sum_{i=1}^{128} \sum_{j=1}^{128} X_{ij}}{128 \times 128} \right) \bigg/ \left(1 - \frac{\sum_{i=1}^{128} \sum_{j=1}^{128} X_{ij}}{128 \times 128}\right) =
$$
$$
= \frac{\sum_{i=1}^{128} \sum_{j=1}^{128} X_{ij}}{128 \times 128 - \sum_{i=1}^{128} \sum_{j=1}^{128} X_{ij}}
$$

Due to the procedure for constructing reference data, the ratio of a priori probabilities of hypotheses coincides with the partition threshold, i.e., $\frac{P(H_1)}{P(H_0)}=p$. Thus, the space of all possible masks is divided into two disjoint regions: the critical region of hypothesis $H_0$ and the critical region of hypothesis $H_1$.

\begin{figure}
  \centering
  \includegraphics[width=.8\linewidth]{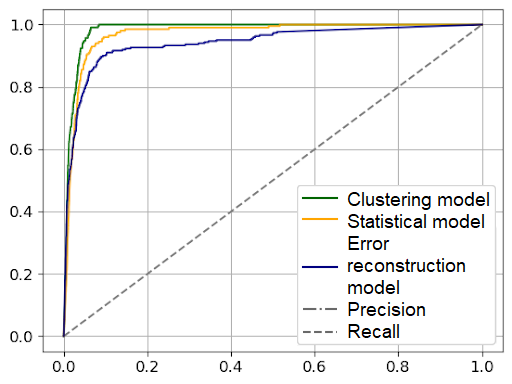}
  \caption{ROC Curves of the Classifier Relative to Model Predictions}
  \label{fig:17}
\end{figure}

The constructed error curve then evaluates the quality of the classifier (\ref{eq:16}) relative to the predictions of each model. Figure \ref{fig:17} shows the ROC curves and their corresponding AUC indicators. It can be seen that the UT78 mask corresponds quite accurately to the  foreground outliers highlighted by each model. For the DBSCAN model, the maximum specificity of the classifier (\ref{eq:16}) is achieved with the lowest sensitivity. These observations correspond to the accuracy indicators (\ref{eq:14}) in Figure \ref{fig:16}.

\section{Map of Foreground Outliers}

For a detailed analysis of the results, two sections of the map were selected for in-depth study. The error reconstruction method was chosen as the primary model because it provides additional information about the pixels of the reconstructed image, which can help interpret the results. The first section has a right ascension of 00h00m00s -- 02h00m00s and a declination of -40° -- -70° in the equatorial coordinate system. According to the HyberLEDA database, this section contains a single bright star -- HD009992 \cite{51}. The second section has a right ascension of 12h20m00s -- 13h40m00s and a declination of -45° -- -65° in the equatorial coordinate system. This section does not contain any known objects with a brightness of 1 magnitude or brighter, according to the HyberLEDA database \cite{51}.

Each section was projected onto a plane. Then, a sample was compiled, including random fragments of the specified area measuring 128×128 pixels. As a result of applying the error reconstruction model, the reconstruction error $E_r(X)$ was calculated for each image in the sample. Additionally, for more detailed information, the maximum error was calculated for each image in the sample using the formula:

$$
E_{max}(X) = \max_{i,j=1 \cdots 128} \left\{ 
\left[ vec(X) - vec \left( g_{\hat{\theta}_2} \left( f_{\hat{\theta}_1} \left( X
\right) \right) \right) \right]^2_{ij}
\right\},
$$
where $vec$ represents the transformation of the image into a vector, and squaring occurs coordinate-wise.

 The training set was built by extracting many small patches from the data, after applying a rough galactic cut. This cut leaves a lot of diffuse foreground contamination. As is evident from the very non-Gaussian histograms in fig. \ref{fig:3}, especially at high frequencies, which are known to be characterized by massive dust contamination. These foreground outliers are strongly anisotropic. This means that our test dataset contains predominantly isotropic patches. Moreover, we computed the absolute error $L_1$ of the model predictions. The error distribution on the test sample approximates a normal distribution with the mean $\mu=0.3$, but deviates due to the non-negativity constraint on the errors.

 Given our understanding of the error distribution within the test sample, we anticipate that the distribution of $E_{r}(X)$ in the specified sections will closely follow a chi-squared distribution if there are no foreground outliers in the test sections. In the presence of foreground outliers, the patches will exhibit a high degree of non-stationarity. Additionally, the decoder's training on anomalous patches containing foreground outliers was suboptimal. These two factors will lead to the decoder not being able to qualitatively restore anomalous patches that contain foreground outliers. Then the error $E_{r}(X)$ on these patches will be higher.

 This will distort the entire distribution towards an increase in the number of patches with higher errors. Also, if the foreground outliers is localized in one region of the section, the patch sample can be considered as two distinct samples from different distributions. Consequently, the overall $E_{r}(X)$ distribution may exhibit bimodal characteristics.

\begin{figure}
\centering
\includegraphics[width=1\linewidth]{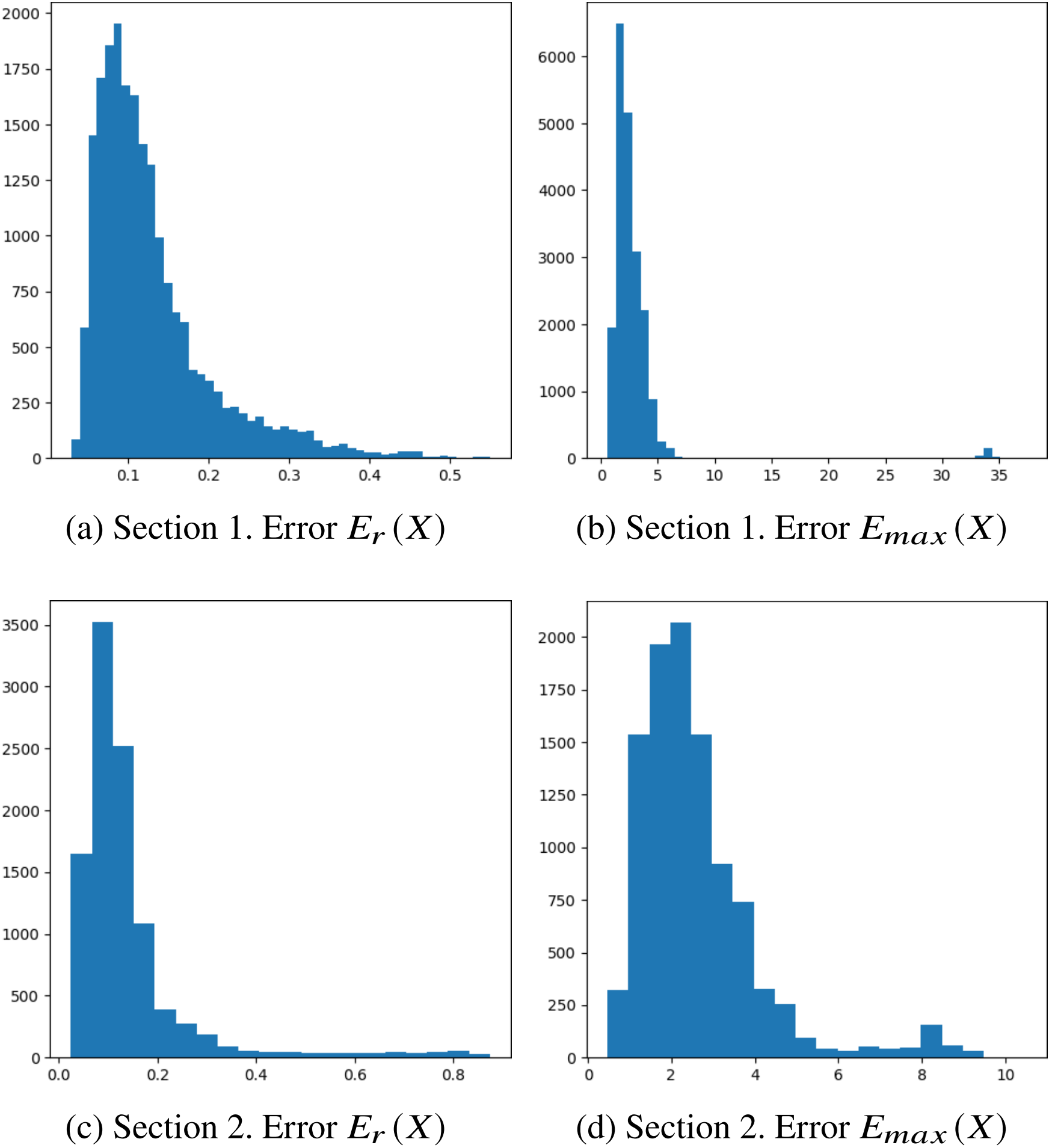}
\caption{Error Distribution Histograms}
\label{fig:19}
\end{figure}

The error distributions were obtained for the first sample area, as shown in Fig. \ref{fig:19}a and Fig. \ref{fig:19}b. The error distributions for the second sample area are shown in Fig. \ref{fig:19}c and Fig. \ref{fig:19}d. It can be seen from the figures that the $E_{max}(X)$ error makes it easier to visually identify abnormal areas, which simplifies the analysis. It is also evident from the error histograms that the first section contains an  foreground outlier, while the second section requires further investigation.  Note that the distribution in Fig. \ref{fig:19} is the model error distribution and depends on the autoencoder output. It is a dimensionless quantity.


\begin{figure}
\centering
\includegraphics[width=1\linewidth]{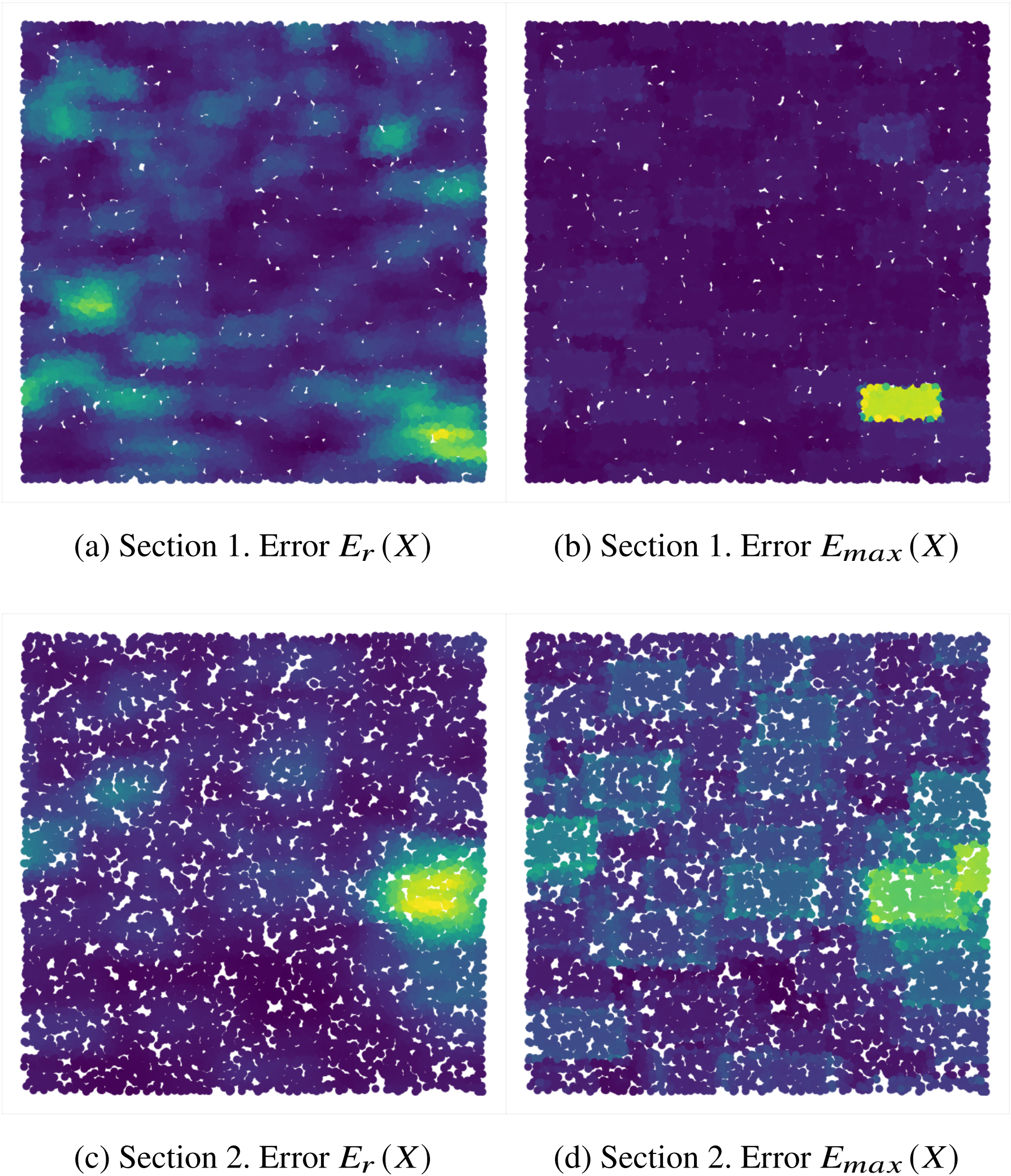}
\caption{Analysis of Reconstruction Errors}
\label{fig:20}
\end{figure}

By associating a point with the center of each image from the sample and assigning an error value to each point, a map of abnormal areas can be created. The map is constructed as follows: the X coordinate of the projected image is plotted horizontally, the Y coordinate is plotted vertically, and the error value is shown in color. The blue color corresponds to the minimum error in the considered distribution, while the yellow color corresponds to the maximum error. Fig. \ref{fig:20} shows the map of abnormal areas for the first and second sections. For both sections, the detected  regions are localized, indicating the presence of atypical objects in these areas of the map.


\begin{figure}
\centering
\includegraphics[width=1\linewidth]{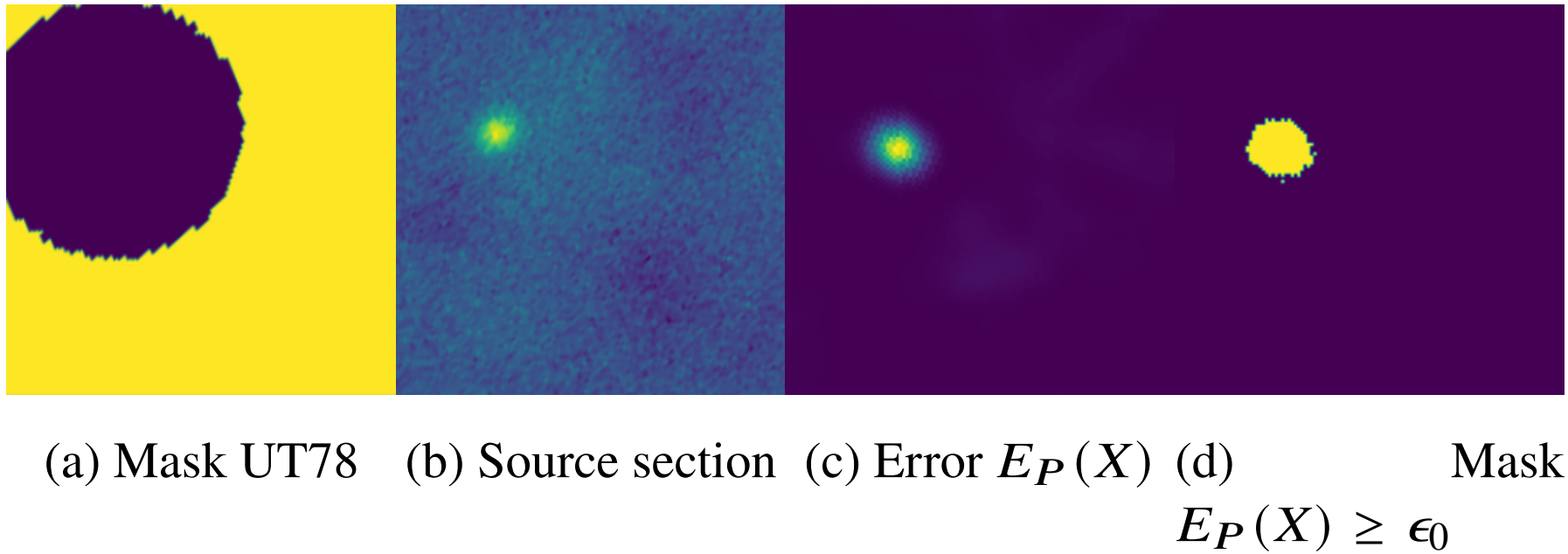}
\caption{Map of Abnormal Areas}
\label{fig:21}
\end{figure}

In all images from the first section with $E_{max}(X)\geq30$, a bright object -- the star HD009992 --was present. Using the error reconstruction method (see Fig. \ref{fig:21}), it is possible to verify that this particular bright object is an  detected foreground outlier. By calculating the pixel error $E_P(X)$ (see Fig. \ref{fig:21}c), a mask $E_P(X) \geq \epsilon_0$ at the level of a single image can be obtained (see Fig. \ref{fig:21}d), where $\epsilon_0 = 30$ is determined from the error distribution histogram \ref{fig:19}. The mask indicates that the object is atypically hot. This behavior is consistent for all images containing the star HD009992.


\begin{figure}
\centering
\includegraphics[width=1\linewidth]{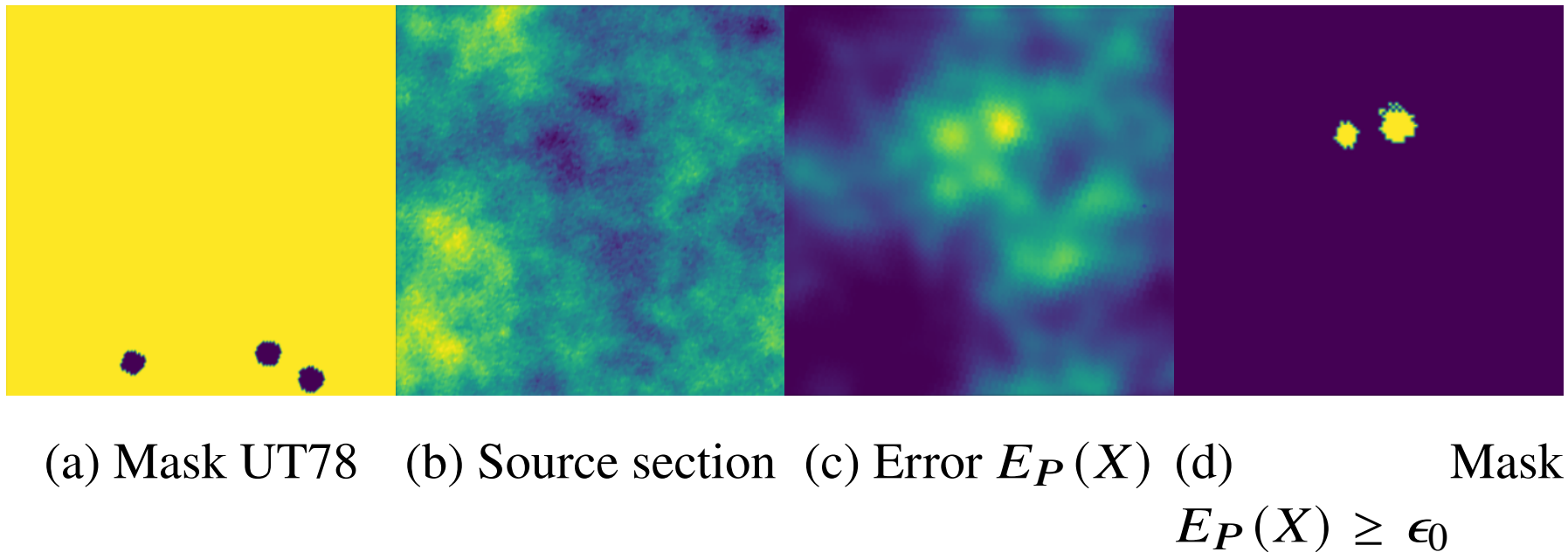}
\caption{Map of Abnormal Areas}
\label{fig:24}
\end{figure}

All images from the second section with $E_{max}(X)>7$ correspond to one area (see Fig. \ref{fig:20}c). However, from the images and the UT78 mask, it is unclear why this area is systematically marked as abnormal. Analysis of reconstruction errors (see Fig. \ref{fig:24}c) for any sample image in the specified area shows that the area contains regions that are atypically cold (see Fig. \ref{fig:24}d). It should be noted that these regions do not correspond to any known object large enough to be detected on maps of relic radiation. This observation may represent model noise, as the autoencoder might not encode such patterns very well. However, the reconstruction error is very small compared to the other  detected regions. This fact does not allow us to unambiguously classify the area as an  foreground outlier, but it does illustrate the possibility of interpreting model results in some ambiguous situations.


By applying the error reconstruction model to all available data, a complete map of  foreground outliers regions can be constructed. Such a map indicates atypical data areas that deviate from the expected pattern. With substantial computing resources, a accurate map can be built, which can complement the mask of  foreground contamination regions in tasks where the accuracy of  abnormal objects detection prevails over completeness. Additionally, constructing a similar map using other data can be extremely useful if building a mask like UT78 is challenging or even impossible with the available data.

\section*{Conclusion}

In this work, an approach using machine learning algorithms was proposed to detect  foreground outliers on astronomical maps, particularly in CMB maps. This approach consists of two stages. In the first stage, features are extracted from the map using a noise-reducing autoencoder with convolutional layers. In the second stage, the extracted features are analyzed using one of the proposed models, and then abnormal areas on the map are identified.

During the analysis of the real maps from the Planck mission, a number of atypical structures were discovered. It was shown that most of the detected objects are indeed  point-like objects or astrophysical contamination regions and can be explained by known astrophysical origin. A small number of  foreground outliers were also found that do not correspond to any known astrophysical  origin. A map of  abnormal areas was compiled, and the quality of the proposed approach was evaluated in comparison with existing models.

 The primary advantage of our model over statistical models is that it does not require prior information about the origin of the targets. This feature allows the model to be used for the discovery of previously unknown objects with unidentified astrophysical properties.

The proposed approach can assist in detecting atypical structures in astronomical maps. However, this approach only identifies the areas of the map that are of the greatest interest. Further research is required to substantiate the astrophysical phenomena that could cause  some of the observed foreground outliers.

\section{Acknowledgements}
We  acknowledge valuable comments of the Referee which helped us to essentially improve the presentation of this work.

\section{Data Availability Statement}

This manuscript has no associated data or the data will not be deposited.

This is theoretical work in which no experimental data are generated and/or
analyzed.

\end{document}